\documentclass[%
 aip,
 jap,%
 amsmath,amssymb,
 reprint,%
]{revtex4-1}

\usepackage[pdftex]{graphicx}
\usepackage{dcolumn}
\usepackage{bm}

\usepackage[caption=false]{subfig}
\usepackage{color}
\usepackage{dcolumn} 
  \newcolumntype{d}[1]{D{.}{.}{#1}}
\usepackage{booktabs}
\usepackage{siunitx}
\usepackage{hyperref}

\begin{document}
\preprint{AIP/123-QED}
\title[Kovacs et.al.]{Numerical optimization of writer and media for bit patterned magnetic recording}
\thanks{The following article appeared in A. Kovacs et al., Journal of Applied Physics {\bf{120}}, 013902 and may be found at \url{http://dx.doi.org/10.1063/1.4954888}. This article may be downloaded for personal use only. Any other use requires prior permissions of the author and AIP Publishing. Copyright (2016) American Institute of Physics.}

\author{A. Kovacs}
 \email{alexander.kovacs@donau-uni.ac.at}
 \affiliation{%
 Center of Integrated Sensor Systems, Danube University Krems, 2700 Wiener Neustadt, Austria
 }
\author{H. Oezelt}
 \affiliation{%
 Center of Integrated Sensor Systems, Danube University Krems, 2700 Wiener Neustadt, Austria
 }
\author{M. E. Schabes}
 \affiliation{%
 Spin Transfer Technologies, 45500 Northport Loop West, Fremont, CA 94538, USA
 }%
\author{T. Schrefl}
 \affiliation{%
 Center of Integrated Sensor Systems, Danube University Krems, 2700 Wiener Neustadt, Austria
 }%

\date{\today}

\begin{abstract}
In this work we present a micromagnetic study of the performance potential of bit-patterned (BP) magnetic recording media via joint optimization of the design of the media and of the magnetic write heads. Because the design space is large and complex, we developed a novel computational framework suitable for parallel implementation on compute clusters. Our technique combines advanced global optimization algorithms and finite-element micromagnetic solvers. Targeting data bit densities of $\SI{4}{Tb/in^2}$, we optimize designs for centered, staggered, and shingled BP writing. The magnetization dynamics of the switching of the exchange-coupled composite BP islands of the media is treated micromagnetically. Our simulation framework takes into account not only the dynamics of on-track errors but also of the thermally induced adjacent-track erasure. With co-optimized write heads, the results show superior performance of shingled BP magnetic recording where we identify two particular designs achieving write bit-error rates of $1.5\mathrm{x}10^{-8}$ and $8.4\mathrm{x}10^{-8}$, respectively. A detailed description of the key design features of these designs is provided and contrasted with centered and staggered BP designs which yielded write bit error rates of only $2.8\mathrm{x}10^{-3}$ (centered design) and $1.7\mathrm{x}10^{-2}$ (staggered design) even under optimized conditions.
\end{abstract}



\maketitle
\section{\label{sec:introduction}Introduction}
Data storage capacity of perpendicular magnetic re-cording has increased significantly and the technology itself dominated the industry over the past decade. Currently the annual increase of areal density has slowed down, due to the so called trilemma of magnetic recording, where thermal stability, write-ability and media signal-to-noise ratio are conflicting requirements\citep{Richter1999,Kryder2005}. By shrinking the bit size the areal density is increased but also the risk of spontaneous thermally induced reversal. To ensure thermal stability magnetically harder materials, i.e. with higher magnetocrystalline anisotropy, can be used. This leads to higher switching field of the media and therefore a higher write field, limited by the size of the recording head, is needed. A bigger write head and a stronger write field broadens the written track, which is counter-productive for high areal densities. 

One promising technology for extending magnetic storage densities towards $\SI{10}{Tb/in^2}$ is bit patterned magnetic recording and many papers have been published showing its potential\citep{wood2000,richter2006a,schabes2008,muraoka2011a,dong2012}. Recording on bit patterned media, where stored information with random clusters of grains (in conventional recording) is replaced by patterned magnetic islands, achieves higher signal-to-noise ratio and thermal stability.
However, bit patterned magnetic recording still pose many novel challenges in terms of media fabrication\citep{Terris2005} but also in terms of recording physics\citep{schabes2008}. A localized write field with high down-track and cross-track field gradients are important for addressability of such high areal density media. The distribution of the write field considerably depends on the pole tip's shape and shield distances.
For best writer performance both the effective write field and write field gradient should be maximized.
To avoid unintended erasure of nearby bits the fringing field has to be localized to the pole of the write head.

These recording field requirements also strongly depend on the media. Cross-track and down-track pitch, shape, pattern, composition and intrinsic material properties of the media drastically change the requirements on the write field distribution. Exchange coupled composite media consists of a soft magnetic layer, acting as nucleation site therefore reducing the media's switching field, and a hard magnetic layer, acting as storage layer increasing the media's thermal stability. 

All these influential decisions in design configuration make it difficult to find the most suitable recording head for a specific application. In other words, finding the best write head - media design ends up to be a high dimensional optimization problem where each new design parameter adds another dimension to the configuration space.
What kind of search strategy can be used depends on the balance between computational power and the model evaluation's grade of simplification.
Following a design of experiment approach, where multiple sweeps of a single design parameter have to be performed, would lead to a numerically too expensive search for the optimal design. Fewer computational expenses can be achieved by using an optimization software environment which intelligently decides which regions of the design space are more promising in order to find the best solution.
Kalezhi et al. showed a statistical media property optimization approach\cite{Kalezhi2012} where the switching probabilities of single phase and exchange coupled composite media are derived through their energy barriers and energy barrier gradients. This method accurately considers the thermal processes when evaluating the error rates. In the method presented in this work thermal processes are taken into account for the error rate on the adjacent track whereas the writing dynamics such as field rise time and head motion is taken into account for evaluating the error rates on the target and previous bit.
Bashir et al. proposed a single objective optimization approach\citep{Bashir2012}. In their work the effective field and effective field gradient are maximized within the geometrical constraints and then only afterwards the magnetocrystalline anisotropy constant of the media is tuned to optimize the switching field of the media.
Fukuda and co-workers\citep{Fukuda2012} simultaneously optimized writer and media parameters for granular perpendicular recording with a multi-objective approach, using a genetic algorithm together with a finite element static Maxwell solver and a micromagnetic solver. A multi-objective optimization software can be used in combination with a response surface method\citep{Kovacs2014} to reduce the number of model evaluations but might involve loss of expressiveness for the optimization's decisions.

Writing on continuous granular media, where a bit is formed by a large group of grains, no loss of information immediately appears if just a few grains are not switched by the write field as long as the transition between two bit cells is still detectable by the reader. But looking at bit patterned media, where each bit cell is formed by just one single island, we now have to assess if switching has occurred or not and introduce bit error rates \citep{muraoka2011a,muraoka2008}. Similarly, adjacent track erasure which influences only a few grains of a bit in conventional recording\citep{Dean2008}. In bit patterned recording a bit error occurs if an island on the adjacent track switches its magnetization in the fringing field of the writer.

Hence, a meaningful objective function is to minimize the total write error of a single island. 
In the case of exchange coupled composite media, the calculation of the effective write field is not as trivial as for single phase media islands. So in this work we present a total write error rate calculation which is fully computed with micro-magnetic simulations. We minimize the total bit error rate with a single objective optimization for three different recording schemes: centered writing, staggered writing and shingled writing.

The paper is organized as follows: In Section~\ref{sec:method} we describe the sequence of the optimization cycle, which parameters of the write head and media are optimized and how we calculate the total write error rate. In Section~\ref{sec:results} the two best write head - media designs for each writing scheme and their error rates are presented and discussed.

\section{\label{sec:method}Method}
In this section we illustrate the iterative optimization cycle and how the models and meshes are generated. We explain the global optimization algorithm and how it makes decisions. We illustrate all design parameters of the recording head and media. Moreover, we show which of them are handled by the optimization algorithm and which are not. Finally we explain how each proposed design is evaluated with its total write error rate $\mathrm{BER}_{\mathrm{tot}}$.
%
\subsection{\label{sec:opt-cycle}Optimization cycle}
\begin{figure}[htb]
\tiny
\begingroup%
  \makeatletter%
  \providecommand\color[2][]{%
    \errmessage{(Inkscape) Color is used for the text in Inkscape, but the package 'color.sty' is not loaded}%
    \renewcommand\color[2][]{}%
  }%
  \providecommand\transparent[1]{%
    \errmessage{(Inkscape) Transparency is used (non-zero) for the text in Inkscape, but the package 'transparent.sty' is not loaded}%
    \renewcommand\transparent[1]{}%
  }%
  \providecommand\rotatebox[2]{#2}%
  \ifx\svgwidth\undefined%
    \setlength{\unitlength}{216.85214844bp}%
    \ifx\svgscale\undefined%
      \relax%
    \else%
      \setlength{\unitlength}{\unitlength * \real{\svgscale}}%
    \fi%
  \else%
    \setlength{\unitlength}{\svgwidth}%
  \fi%
  \global\let\svgwidth\undefined%
  \global\let\svgscale\undefined%
  \makeatother%
  \begin{picture}(1,0.79414922)%
    \put(0,0){\includegraphics[width=\unitlength]{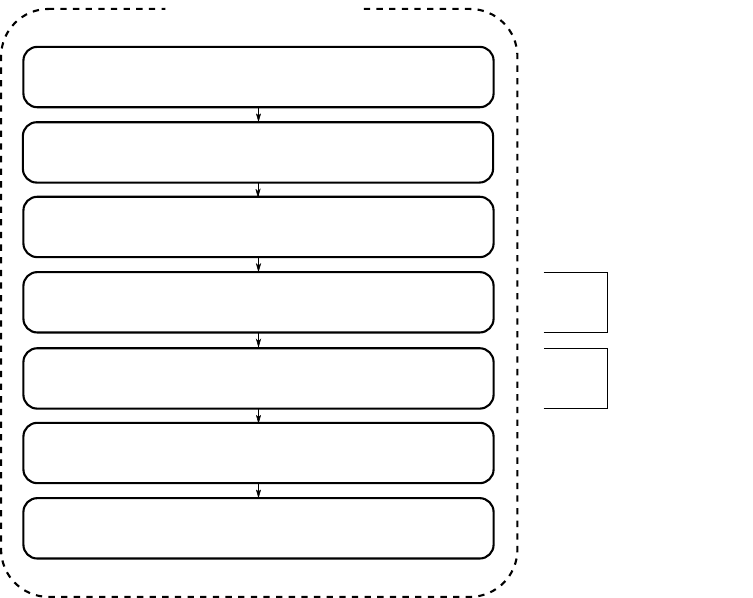}}%
    \put(0.34431125,0.77802682){\color[rgb]{0,0,0}\makebox(0,0)[b]{\smash{{\textit{Model evaluation}}}}}%
    \put(0.34275037,0.70156009){\color[rgb]{0,0,0}\makebox(0,0)[b]{\smash{Prepares the simulation environment}}}%
    \put(0.34332166,0.67356969){\color[rgb]{0,0,0}\makebox(0,0)[b]{\smash{(Python)}}}%
    \put(0.34284741,0.60149444){\color[rgb]{0,0,0}\makebox(0,0)[b]{\smash{Creating geometry and mesh}}}%
    \put(0.3426219,0.57292522){\color[rgb]{0,0,0}\makebox(0,0)[b]{\smash{(Python + SALOME)}}}%
    \put(0.34352838,0.50269263){\color[rgb]{0,0,0}\makebox(0,0)[b]{\smash{Calculation of write head field $H_{\mathrm{head}}$}}}%
    \put(0.34332166,0.47390919){\color[rgb]{0,0,0}\makebox(0,0)[b]{\smash{(FEMME)}}}%
    \put(0.34352838,0.40262695){\color[rgb]{0,0,0}\makebox(0,0)[b]{\smash{Calculation of write error rate $\mathrm{BER}_{\mathrm{targ}}$}}}%
    \put(0.34431125,0.37359444){\color[rgb]{0,0,0}\makebox(0,0)[b]{\smash{(Python + FEMME)}}}%
    \put(0.34352842,0.30151163){\color[rgb]{0,0,0}\makebox(0,0)[b]{\smash{Calculation of write error rate $\mathrm{BER}_{\mathrm{prev}}$}}}%
    \put(0.3433217,0.273307){\color[rgb]{0,0,0}\makebox(0,0)[b]{\smash{(Python + FEMME)}}}%
    \put(0.34352838,0.2024956){\color[rgb]{0,0,0}\makebox(0,0)[b]{\smash{Calculation of write error rate $\mathrm{BER}_{\mathrm{adj}}$}}}%
    \put(0.34332166,0.17371216){\color[rgb]{0,0,0}\makebox(0,0)[b]{\smash{(FEMME)}}}%
    \put(0.34352838,0.10242993){\color[rgb]{0,0,0}\makebox(0,0)[b]{\smash{Calculating and returning $\mathrm{BER}_{\mathrm{tot}}$}}}%
    \put(0.34332166,0.07422529){\color[rgb]{0,0,0}\makebox(0,0)[b]{\smash{(Python)}}}%
    \put(0.83451434,0.40666424){\color[rgb]{0,0,0}\makebox(0,0)[lb]{\smash{12 parallel}}}%
    \put(0.83451434,0.37867386){\color[rgb]{0,0,0}\makebox(0,0)[lb]{\smash{recording simulations}}}%
    \put(0.83381452,0.30554453){\color[rgb]{0,0,0}\makebox(0,0)[lb]{\smash{12 parallel}}}%
    \put(0.83451434,0.27755414){\color[rgb]{0,0,0}\makebox(0,0)[lb]{\smash{recording simulations}}}%
  \end{picture}%
\endgroup%
\caption{\label{fig:flow-chart}Flow chart of a single model evaluation which returns the total error rate $\mathrm{BER}_{\mathrm{tot}}$ after sequentially evaluating $\mathrm{BER}_{\mathrm{targ}}$, $\mathrm{BER}_{\mathrm{prev}}$ and $\mathrm{BER}_{\mathrm{adj}}$. The calculation of $\mathrm{BER}_{\mathrm{targ}}$ and $\mathrm{BER}_{\mathrm{prev}}$ consists of 12 simultaneously performed micro magnetic recording simulations, which search for a critical interaction field $H_{\mathrm{crit}}$ where successful writing of the target bit isn't achievable any more. If the head field is too weak to switch the target dot, the following error rate calculations are aborted and a total error rate of 1 is returned instead.}
\end{figure}

The optimization cycle consists of two major parts. Firstly the {\textit{optimization process}}, which iteratively varies several input parameters of a black box system to minimize an output value of that same system. The second part is located inside this black box. The {\textit{model evaluation}} (FIG.~\ref{fig:flow-chart}) mostly has to be implemented from scratch to read in the given input parameters and return an output value which describes the performance of a given set of input parameters. The optimization algorithm minimizes such an output value by iterating through several combinations of design parameters in a reasonably small amount of model evaluations.

In this work the used algorithm is called \textbf{e}fficient \textbf{g}lobal \textbf{o}ptimization\citep{Jones1998} and comes with the open source optimization software tool 
dakota\citep{Adams2015}. This algorithm is suggested for optimizations of problems with long model evaluation times.
The algorithm, as most optimization algorithms do, fills the design space with initial training points and evaluates them. It is suggested to train the optimization's predictor with at least ten times the amount of design parameters. A nearly uniform distribution of these points throughout this high dimensional design space is achieved by using an optimized Latin hypercube space-filling method\citep{Damblin2013,Park1994}, which is embedded in the used optimization software\citep{Jones1998}. Similar to a response surface method the results of these initial simulations are described with an approximation function, in this case a Gaussian process. In addition to objective function values and numerical derivatives of the objective function the method uses variance data of each already evaluated design and uses it as indicator for future placement of training points. With this information the software calculates an estimated improvement for all unknown design sets, which determines the next training point. The algorithm always evaluates the design set with the highest expected improvement. After evaluating the most promising new design set, all training points are described again with an updated Gaussian process and the expected improvement function is calculated again. This will be iterated until the global optimum is found, or in other words no improvement can be expected.

All proposed designs are characterized via the second part of the optimization cycle, the {\textit{model evaluation}}, which in our case constructs and analyses a write head - media model. The necessary steps of the model evaluation are depicted in FIG.~\ref{fig:flow-chart}. 
The evaluation of a model is performed with a Python script, which starts reading in given design parameters from an input file produced by the {\textit{optimization process}} and constructs a write head geometry accordingly. Computer aided design is done with the software package SALOME\citep{salome}. Meshing with a $\SI{2.5}{nm}$ fine mesh near the pole tip is done with the mesh generation program NETGEN\citep{netgen}.
After these preprocessing steps a hybrid finite element/boundary element method\citep{schrefl2005} is solving the Landau-Lifshitz-Gilbert equation at $\SI{0}{K}$ temperature and calculates the emergent magnetostatic field $\bm{H}_{\mathrm{head}}$ from the recording head between the air bearing surface and the soft under-layer. Starting from a remanent state, a current pulse with a rise time of $\SI{0.1}{ns}$ is applied to the write head.
After $\SI{2}{ns}$ we compute the write field below the saturated write head with a resolution of $\SI{2.5}{nm}$. This 3-dimensional write field data $\bm{H}_{\mathrm{head}}$ is handed over to the total error rate calculation unit in the evaluation script. Due to the decoupled work flow of the write field computation and the actual recording simulations interactions from the media onto the pole and underlayer are not included. Recording is simulated by moving the precomputed write field over the bit with a rise time of $\SI{0.1}{ns}$. An overshoot of coil current has not taken into consideration in this work. The head velocity was $\SI{13.5}{\meter\per\second}$.  All design related input parameters are detailed in Section~\ref{sec:design-params}. How the total write error rate $\mathrm{BER}_{\mathrm{tot}}$ (black box output) is calculated is described in Section~\ref{sec:error-calc}.
\subsection{\label{sec:design-params}Design parameters}
\begin{figure*}[htb]
\begingroup%
  \makeatletter%
  \providecommand\color[2][]{%
    \errmessage{(Inkscape) Color is used for the text in Inkscape, but the package 'color.sty' is not loaded}%
    \renewcommand\color[2][]{}%
  }%
  \providecommand\transparent[1]{%
    \errmessage{(Inkscape) Transparency is used (non-zero) for the text in Inkscape, but the package 'transparent.sty' is not loaded}%
    \renewcommand\transparent[1]{}%
  }%
  \providecommand\rotatebox[2]{#2}%
  \ifx\svgwidth\undefined%
    \setlength{\unitlength}{408.37290039bp}%
    \ifx\svgscale\undefined%
      \relax%
    \else%
      \setlength{\unitlength}{\unitlength * \real{\svgscale}}%
    \fi%
  \else%
    \setlength{\unitlength}{\svgwidth}%
  \fi%
  \global\let\svgwidth\undefined%
  \global\let\svgscale\undefined%
  \makeatother%
  \begin{picture}(1,0.61600935)%
    \put(0,0){\includegraphics[width=\unitlength]{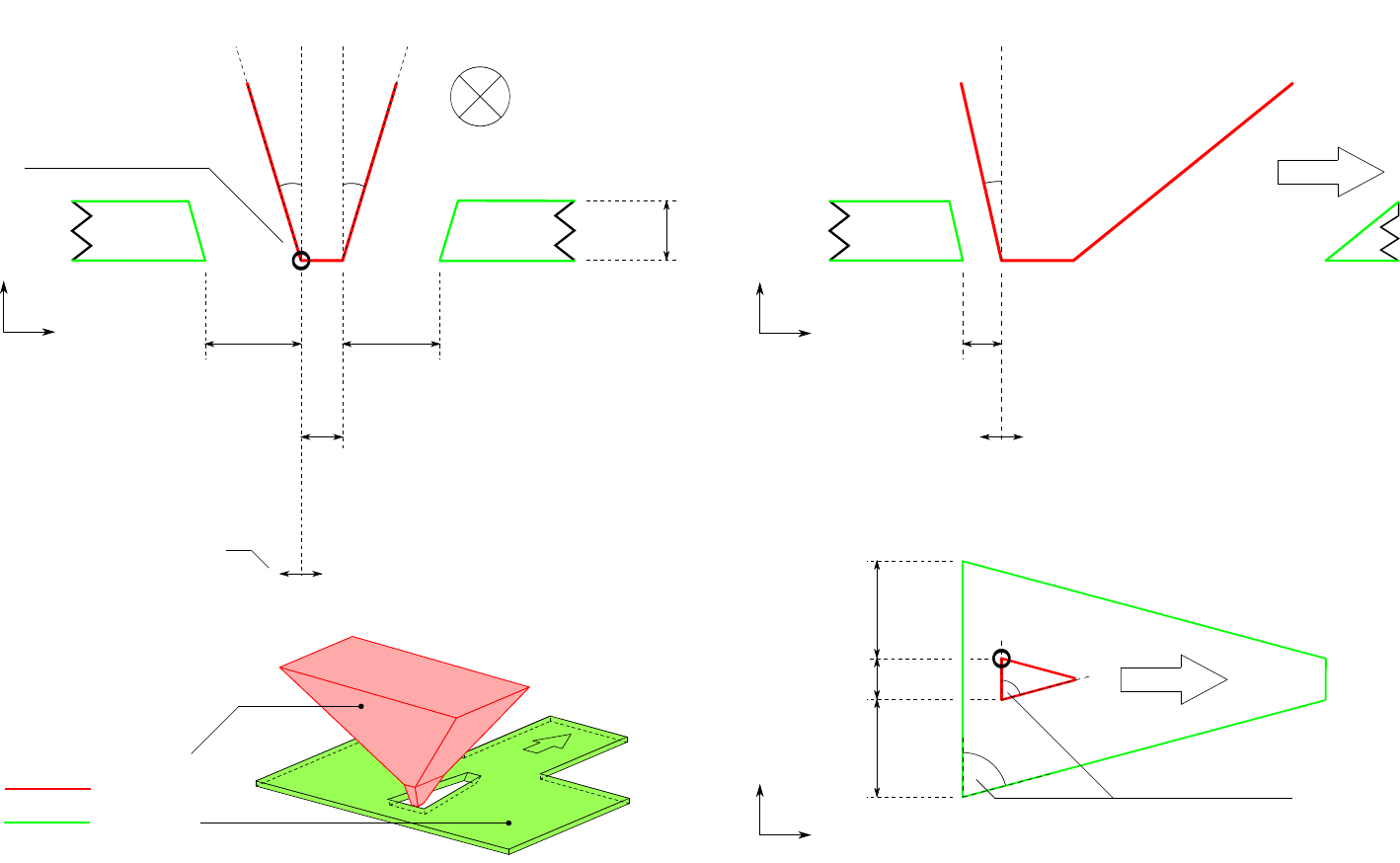}}%
    \put(0.27942539,0.34472822){\color[rgb]{0,0,0}\makebox(0,0)[lb]{\smash{side shield gap}}}%
    \put(0.25551429,0.27827835){\color[rgb]{0,0,0}\makebox(0,0)[lb]{\smash{pole tip width}}}%
    \put(0.22988499,0.59820557){\color[rgb]{0,0,0}\makebox(0,0)[b]{\smash{side edge angle}}}%
    \put(0.47612779,0.49820315){\color[rgb]{0,0,0}\makebox(0,0)[rb]{\smash{shield thickness}}}%
    \put(0.34299704,0.58123369){\color[rgb]{0,0,0}\makebox(0,0)[lb]{\smash{flying direction}}}%
    \put(0.03864783,0.36181426){\color[rgb]{0,0,0}\makebox(0,0)[lb]{\smash{$y$}}}%
    \put(0.01121293,0.41477426){\color[rgb]{0,0,0}\makebox(0,0)[lb]{\smash{$z$}}}%
    \put(0.18082789,0.34472822){\color[rgb]{0,0,0}\makebox(0,0)[rb]{\smash{side shield gap}}}%
    \put(0.00288347,0.53376826){\color[rgb]{0,0,0}\makebox(0,0)[lb]{\smash{write corner}}}%
    \put(0.16999387,0.1856518){\color[rgb]{0,0,0}\makebox(0,0)[rb]{\smash{cross track offset}}}%
    \put(0.13640402,0.2203082){\color[rgb]{0,0,0}\makebox(0,0)[rb]{\smash{(shingled)}}}%
    \put(0.00288347,0.50813873){\color[rgb]{0,0,0}\makebox(0,0)[lb]{\smash{(shingled)}}}%
    \put(0.70155298,0.34475219){\color[rgb]{0,0,0}\makebox(0,0)[rb]{\smash{trailing shield gap}}}%
    \put(0.75370722,0.59820661){\color[rgb]{0,0,0}\makebox(0,0)[rb]{\smash{trailing edge angle}}}%
    \put(0.71538935,0.27827773){\color[rgb]{0,0,0}\makebox(0,0)[rb]{\smash{down track offset}}}%
    \put(0.91274016,0.52098462){\color[rgb]{0,0,0}\makebox(0,0)[lb]{\smash{flying direction}}}%
    \put(0.57880089,0.36080507){\color[rgb]{0,0,0}\makebox(0,0)[lb]{\smash{$x$}}}%
    \put(0.55136599,0.41376507){\color[rgb]{0,0,0}\makebox(0,0)[lb]{\smash{$z$}}}%
    \put(0.82940828,0.05500193){\color[rgb]{0,0,0}\makebox(0,0)[lb]{\smash{pole tip taper angle}}}%
    \put(0.80067673,0.1584163){\color[rgb]{0,0,0}\makebox(0,0)[lb]{\smash{flying}}}%
    \put(0.54382504,0.12661832){\color[rgb]{0,0,0}\makebox(0,0)[b]{\smash{pole tip width}}}%
    \put(0.53738427,0.07710739){\color[rgb]{0,0,0}\makebox(0,0)[b]{\smash{side shield gap}}}%
    \put(0.53738432,0.17612926){\color[rgb]{0,0,0}\makebox(0,0)[b]{\smash{side shield gap}}}%
    \put(0.57880089,0.00251121){\color[rgb]{0,0,0}\makebox(0,0)[lb]{\smash{$x$}}}%
    \put(0.55136599,0.05547121){\color[rgb]{0,0,0}\makebox(0,0)[lb]{\smash{$y$}}}%
    \put(0.80067673,0.09663032){\color[rgb]{0,0,0}\makebox(0,0)[lb]{\smash{direction}}}%
    \put(0.07701045,0.04762303){\color[rgb]{0,0,0}\makebox(0,0)[lb]{\smash{main pole}}}%
    \put(0.07701045,0.02199364){\color[rgb]{0,0,0}\makebox(0,0)[lb]{\smash{shield}}}%
    \put(0.00207567,0.60887865){\color[rgb]{0,0,0}\makebox(0,0)[lb]{\smash{(a)}}}%
    \put(0.5421391,0.60883752){\color[rgb]{0,0,0}\makebox(0,0)[lb]{\smash{(b)}}}%
    \put(0.54222877,0.216549){\color[rgb]{0,0,0}\makebox(0,0)[lb]{\smash{(c)}}}%
  \end{picture}%
\endgroup%
\caption{\label{fig:main-pole-design-parameters} (color-online) The main pole (red), wrap around shield (green) and their free parameters. The (a) down track view of pole tip and shield with the writing scheme dependent cross track offset, (b) cross track view and (c) top view. The pole tip taper angle is always kept at $\SI{75}{\degree}$.}
\end{figure*}
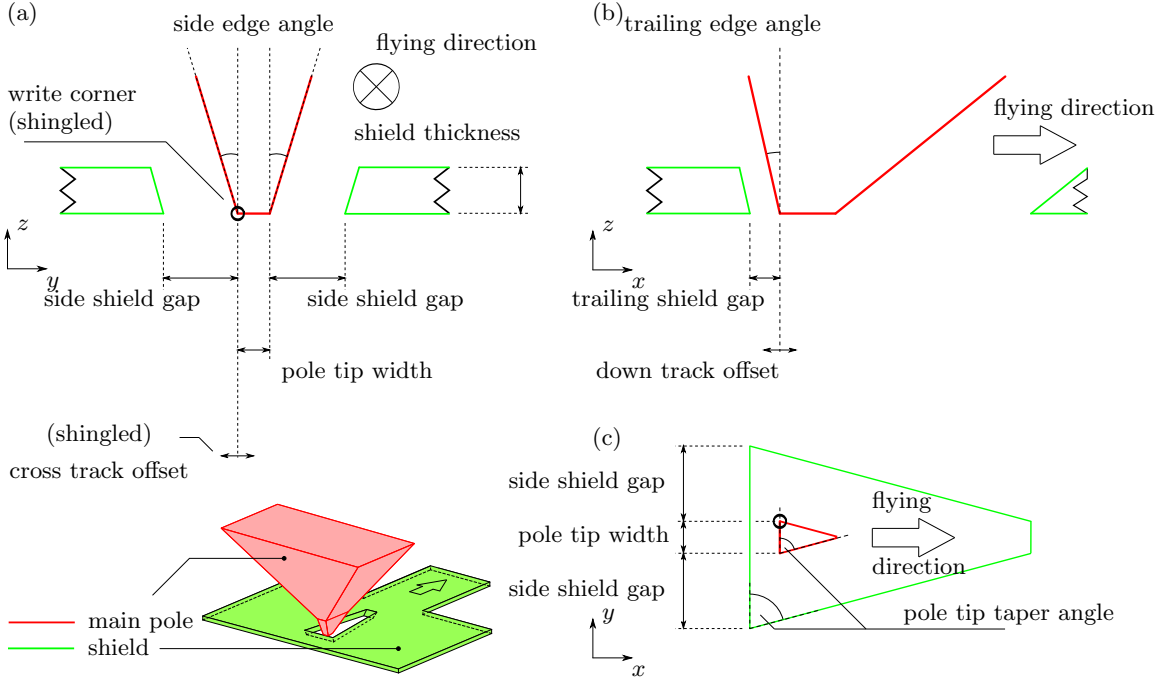

In FIG.~\ref{fig:main-pole-design-parameters} the free design parameters of the write head and the contour of a wrap around shield are shown. In FIG.~\ref{fig:main-pole-design-parameters}a the main pole and shield is pictured in down track direction, where one can see the side gap of the shield, the shield thickness, the side edge angles, the pole tip width and the cross track offset parameter. FIG.~\ref{fig:main-pole-design-parameters}b represents a cross section of the geometry in cross track direction illustrating the trailing shield gap, trailing edge angle and the down track offset parameter. In FIG.~\ref{fig:main-pole-design-parameters}c the write head and wrap around shield is shown from the media's perspective looking up to the air bearing surface, depicting the definition of the pole tip taper angle.

For each optimization of a specific writing scheme (centered, staggered and shingled), we optimized the trailing shield gap, trailing edge angle, side shield gap, side edge angle and trailing edge down track position. All described angle definitions correspond to the main pole's faces. The wrap around shield faces are constructed parallel to their opposite pole tip face. Considering skewing, the taper angle from trailing edge to leading edge of the write head's air bearing surface (see FIG.~\ref{fig:main-pole-design-parameters}c) is kept constant at $\SI{75}{\degree}$ during optimization which results in non-planar pole tip faces in cross track orientation. As for centered and staggered writing we assume a cross track offset of $\SI{0}{\nano\meter}$ as optimal cross track position of the head due to the symmetry of the writing schemes and include the shield thickness as free parameter for the optimization process. Whereas for shingled recording we additionally include the cross track offset as free parameter to get optimized. The pole tip width and the shield thickness at $\SI{80}{\nano\meter}$ and $\SI{50}{\nano\meter}$ respectively are kept constant for this scheme.

\begin{figure}[htb]
\subfloat[top-down view of media layout\label{sfig:media-layout-top}]{%
\begingroup%
  \makeatletter%
  \providecommand\color[2][]{%
    \errmessage{(Inkscape) Color is used for the text in Inkscape, but the package 'color.sty' is not loaded}%
    \renewcommand\color[2][]{}%
  }%
  \providecommand\transparent[1]{%
    \errmessage{(Inkscape) Transparency is used (non-zero) for the text in Inkscape, but the package 'transparent.sty' is not loaded}%
    \renewcommand\transparent[1]{}%
  }%
  \providecommand\rotatebox[2]{#2}%
  \ifx\svgwidth\undefined%
    \setlength{\unitlength}{155.5560791bp}%
    \ifx\svgscale\undefined%
      \relax%
    \else%
      \setlength{\unitlength}{\unitlength * \real{\svgscale}}%
    \fi%
  \else%
    \setlength{\unitlength}{\svgwidth}
  \fi%
  \global\let\svgwidth\undefined%
  \global\let\svgscale\undefined%
  \makeatother%
  \begin{picture}(1,0.97389563)%
    \put(0,0){\includegraphics[width=\unitlength]{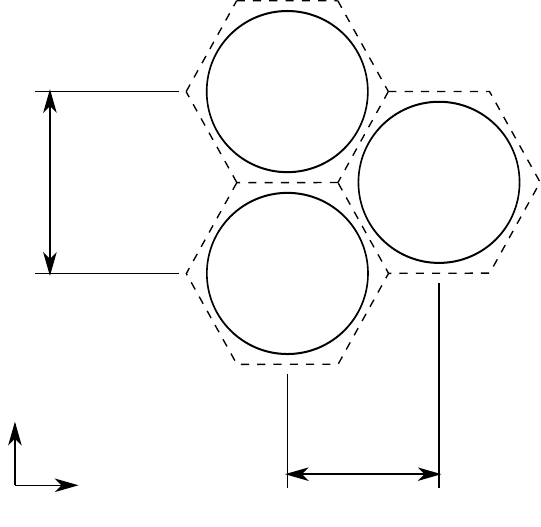}}%
    \put(0.03188173,0.21591705){\color[rgb]{0,0,0}\makebox(0,0)[lb]{\smash{$x$}}}%
    \put(0.1659391,0.07542688){\color[rgb]{0,0,0}\makebox(0,0)[lb]{\smash{$y$}}}%
    \put(0.53102124,0.79325108){\color[rgb]{0,0,0}\makebox(0,0)[b]{\smash{previous}}}%
    \put(0.81205522,0.62520477){\color[rgb]{0,0,0}\makebox(0,0)[b]{\smash{adjacent}}}%
    \put(0.53212582,0.45759132){\color[rgb]{0,0,0}\makebox(0,0)[b]{\smash{target}}}%
    \put(0.40339986,0.00836753){\color[rgb]{0,0,0}\makebox(0,0)[lb]{\smash{cross track pitch}}}%
    \put(0.33468189,0.83123979){\color[rgb]{0,0,0}\makebox(0,0)[rb]{\smash{down track pitch}}}%
    \put(0.67247678,0.12901113){\color[rgb]{0,0,0}\makebox(0,0)[b]{\smash{$\SI{11.7}{nm}$}}}%
    \put(0.11758938,0.62091384){\color[rgb]{0,0,0}\makebox(0,0)[lb]{\smash{$\SI{13.5}{nm}$}}}%
    \put(0.34612037,0.33987007){\color[rgb]{0,0,0}\makebox(0,0)[rb]{\smash{$d_{\mathrm{dot}}=\SI{12}{nm}$}}}%
  \end{picture}%
\endgroup%
}\hfill
\subfloat[cross section of one cylindrical media island\label{sfig:media-layout-cross}]{%
\begingroup%
  \makeatletter%
  \providecommand\color[2][]{%
    \errmessage{(Inkscape) Color is used for the text in Inkscape, but the package 'color.sty' is not loaded}%
    \renewcommand\color[2][]{}%
  }%
  \providecommand\transparent[1]{%
    \errmessage{(Inkscape) Transparency is used (non-zero) for the text in Inkscape, but the package 'transparent.sty' is not loaded}%
    \renewcommand\transparent[1]{}%
  }%
  \providecommand\rotatebox[2]{#2}%
  \ifx\svgwidth\undefined%
    \setlength{\unitlength}{216.85039063bp}%
    \ifx\svgscale\undefined%
      \relax%
    \else%
      \setlength{\unitlength}{\unitlength * \real{\svgscale}}%
    \fi%
  \else%
    \setlength{\unitlength}{\svgwidth}%
  \fi%
  \global\let\svgwidth\undefined%
  \global\let\svgscale\undefined%
  \makeatother%
  \begin{picture}(1,0.49698695)%
    \put(0,0){\includegraphics[width=\unitlength]{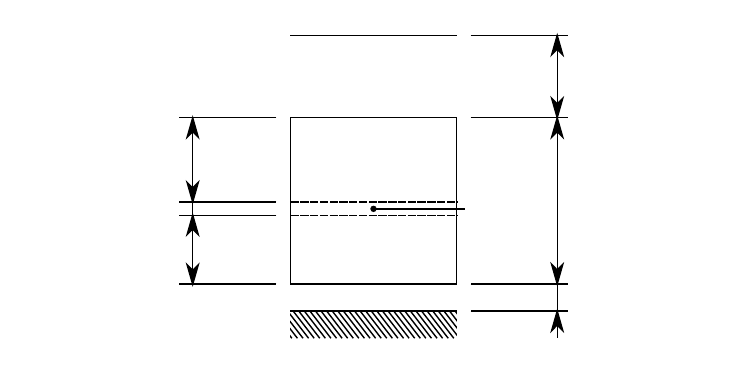}}%
    \put(0.39565672,0.47444594){\color[rgb]{0,0,0}\makebox(0,0)[lb]{\smash{air bearing surface}}}%
    \put(0.39191052,0.00617123){\color[rgb]{0,0,0}\makebox(0,0)[lb]{\smash{soft under layer}}}%
    \put(0.76599943,0.38419466){\color[rgb]{0,0,0}\makebox(0,0)[lb]{\smash{$\SI{6}{nm}$}}}%
    \put(0.76599943,0.21914839){\color[rgb]{0,0,0}\makebox(0,0)[lb]{\smash{$\SI{12}{nm}$}}}%
    \put(0.76599943,0.09075641){\color[rgb]{0,0,0}\makebox(0,0)[lb]{\smash{$\SI{2}{nm}$}}}%
    \put(0.44085778,0.15564464){\color[rgb]{0,0,0}\makebox(0,0)[lb]{\smash{$K_{\mathrm{1,hard}}$}}}%
    \put(0.21512465,0.27359072){\color[rgb]{0,0,0}\makebox(0,0)[rb]{\smash{$\SI{6}{nm}$}}}%
    \put(0.21512465,0.15423277){\color[rgb]{0,0,0}\makebox(0,0)[rb]{\smash{$\SI{5}{nm}$}}}%
    \put(0.21512465,0.21171523){\color[rgb]{0,0,0}\makebox(0,0)[rb]{\smash{$\SI{1}{nm}$}}}%
    \put(0.62110631,0.2131271){\color[rgb]{0,0,0}\makebox(0,0)[lb]{\smash{$A_{\mathrm{ex,int}}$}}}%
    \put(0.44085778,0.26690102){\color[rgb]{0,0,0}\makebox(0,0)[lb]{\smash{$K_{\mathrm{1,soft}}$}}}%
  \end{picture}%
\endgroup%
}
\caption{Media layout \protect\subref{sfig:media-layout-top} top-down view and \protect\subref{sfig:media-layout-cross} cross section of $\SI{4}{Tb/in^2}$ exchange coupled composite media with a dot diameter of $d_{\mathrm{dot}}=\SI{12}{nm}$. The media is pseudo-hexagonally aligned with a $\SI{11.7}{nm}$ cross track pitch and $\SI{13.5}{nm}$ down track pitch. $K_{\mathrm{1,hard}}$ and $A_{\mathrm{ex,int}}$ are included as design parameters into the optimization procedure. Soft phase anisotropy is one fifth of hard phase anisotropy.}
\label{fig:media-layout}
\end{figure}

In FIG.~\ref{sfig:media-layout-top} one can see the pseudo hexagonal media layout with a down track pitch of $\SI{13.5}{nm}$, cross track pitch of $\SI{11.7}{nm}$, a dot diameter of $\SI{12}{nm}$, a bit height of $\SI{12}{nm}$ and a mag-spacing of $\SI{6}{nm}$. This results in an estimated areal density of $\SI{4.08}{Tb/in^2}$ and a filling factor of $\SI{71.6}{\percent}$. FIG.~\ref{sfig:media-layout-cross} shows the additional free media parameters, hard phase's magneto-crystalline anisotropy $K_{\mathrm{1,hard}}$ and interface exchange constant $A_{\mathrm{ex,int}}$ of an interlayer with a thickness of $\SI{1}{nm}$. The soft phase anisotropy is proportionally to the hard phase anisotropy $K_{\mathrm{1,soft}}=0.2K_{\mathrm{1,hard}}$. The magnetic saturation polarization of both phases is $\SI{0.7}{\tesla}$. In this work we assume perfectly aligned media islands where bit position jitter is not included but will be studied in the future.

TABLE~\ref{tab:free-design-params} summarizes the free parameters used during optimization and their parameter search range defined in the optimization. All design parameter ranges in this work are treated as continuous variables by the optimization algorithm.

\begingroup
\squeezetable
\begin{table}[htb]
\begin{ruledtabular}
\begin{tabular}{c c r r rS[table-format=3.1]}
parameter & unit & \multicolumn{1}{c}{lower bound} & \multicolumn{1}{l}{initial} & \multicolumn{1}{l}{upper bound}\\
\hline
\multicolumn{1}{l}{trailing shield gap} & $\mathrm{\si{\nano\meter}}$ & 5.0 & 9.0 & 20.0 \\
\multicolumn{1}{l}{trailing edge angle} & $\mathrm{\si{\degree}}$ & 5.0 & 20.0 & 45.0 \\
\multicolumn{1}{l}{side shield gap} & $\mathrm{\si{\nano\meter}}$ & 5.0 & 16.0 & 20.0\\
\multicolumn{1}{l}{side edge angle} & $\mathrm{\si{\degree}}$ & 5.0 & 15.0 & 45.0 \\
\multicolumn{1}{l}{pole tip width} & $\mathrm{\si{\nano\meter}}$ & 5.0 & 14.0 & 20.0\\
\multicolumn{1}{l}{pole tip taper angle} & $\mathrm{\si{\degree}}$ & 75.0 & 75.0 & 75.0\\
\multicolumn{1}{l}{shield thickness} & $\mathrm{\si{\nano\meter}}$ & 3.0 & 10.0 & 20.0\\
\multicolumn{1}{l}{$K_{\mathrm{1,hard}}$} &  $\mathrm{\si{\mega\joule/\cubic\meter}}$ & 0.5 & 0.8 & 1.2\\
\multicolumn{1}{l}{$A_{\mathrm{ex,int}}$} & $\mathrm{\si{\pico\joule/\meter}}$ & 1.0 & 2.0 & 10.0 \\
\multicolumn{1}{l}{down track offset} & $\mathrm{\si{\nano\meter}}$ & 0.0 & 6.8 & 8.0\\
\multicolumn{1}{l}{cross track offset} & $\mathrm{\si{\nano\meter}}$ & -8.0 & 0.0 & 8.0
\end{tabular}
\caption{\label{tab:free-design-params}  Initial value, lower and upper bound for each free parameter. All parameters, from the head and media, are treated simultaneously.}
\end{ruledtabular}
\end{table}
\endgroup

The geometry of each write head design proposed by the optimization algorithm is constructed with a Python script controlling the pre-processing computer-aided design software SALOME\citep{salome}. The script has two major options for the write head and shield geometry generation. The first option constructs a model consisting of a full write head structure with coils, yoke, return pole, shield and soft underlayer. The second and more storage conservative option produces a smaller wrap around shield and a main pole which are both mirrored along the soft under-layer's upper surface. With this second option the write head is just meshed up to the yoke's edge located $\SI{153}{nm}$ above the air bearing surface. Above this edge and underneath the mirrored main pole part we apply a charge sheet\citep{Shen2008} with a magnetic charge density equalling to the magnetic saturation polarization $J_\mathrm{S}$ instead of using fully meshed return pole, yoke and soft under-layer. Each write head field is calculated with the charge sheet approach in this work. The charge sheet method reduces the computation time by about a factor of 5. Therefore we used the charge sheet method in order to be able to carry out all optimization on a desktop workstation.

TABLE~\ref{tab:material-properties} lists material parameters of each recording head's part used during simulation for both geometry construction options.

\begingroup
\squeezetable
\begin{table}[htb]
\begin{ruledtabular}
\begin{tabular}{l r r r rS[table-format=2.2]}
& \multicolumn{1}{c}{material} & \multicolumn{1}{c}{$K_{\mathrm{1}} \left(\mathrm{\si{\joule/\cubic\meter}}\right)$} & \multicolumn{1}{c}{$J_{\mathrm{S}} \left(\mathrm{\si{\tesla}}\right)$} & \multicolumn{1}{c}{$A_{\mathrm{ex}} \left(\mathrm{\si{\pico\joule/\meter}}\right)$} \\
\hline
main pole & CoFe & 800 & 2.4 & 20.2 \\
yoke & $\mathrm{Ni}_{\mathrm{45}}\mathrm{Fe}_{\mathrm{55}}$ & 0 & 2.0 & 13.0 \\
return pole & $\mathrm{Ni}_{\mathrm{45}}\mathrm{Fe}_{\mathrm{55}}$ & 0 & 2.0 & 13.0 \\ 
shield & CoFe & 800 & 2.4 & 20.2 \\ 
soft under layer & NiFe & 0 & 1.0 & 13.5
\end{tabular}
\caption{\label{tab:material-properties} Fixed material properties used during simulation. Note that shield and main pole have identical material constants.}
\end{ruledtabular}
\end{table}
\endgroup

\subsection{\label{sec:error-calc}Error rate calculation}
The total bit error rate $\mathrm{BER}_{\mathrm{tot}}$ depends on various design parameters which may include key elements of the writer geometry as well as the media properties. Therefore the total bit error rate $\mathrm{BER}_{\mathrm{tot}}$ is used to quantify the performance of each single design configuration.
The single objective function is to minimize the overall bit error rate which is a sum of the following three error occurrences: 1) not writing the target bit $\mathrm{BER}_{\mathrm{targ}}$, 2) accidentally rewriting the previously written bit $\mathrm{BER}_{\mathrm{prev}}$ and 3) the thermally induced writing of an adjacent bit after an unknown amount of recording field passes $\mathrm{BER}_{\mathrm{adj}}$. In this section we illustrate the calculation of these three error rates.

To obtain $\mathrm{BER}_{\mathrm{targ}}$ and $\mathrm{BER}_{\mathrm{prev}}$ we firstly assume a switching field distribution with a standard deviation $\sigma=0.015H_{\mathrm{A}}$ where $H_{\mathrm{A}}=2K_{\mathrm{1,hard}}/J_{\mathrm{S}}$. For example with a hard phase anisotropy of $K_{\mathrm{1,hard}}=$ \SI{750}{\kilo\joule\per\cubic\meter} the distribution's $\sigma$ is \SI{40}{mT}. 

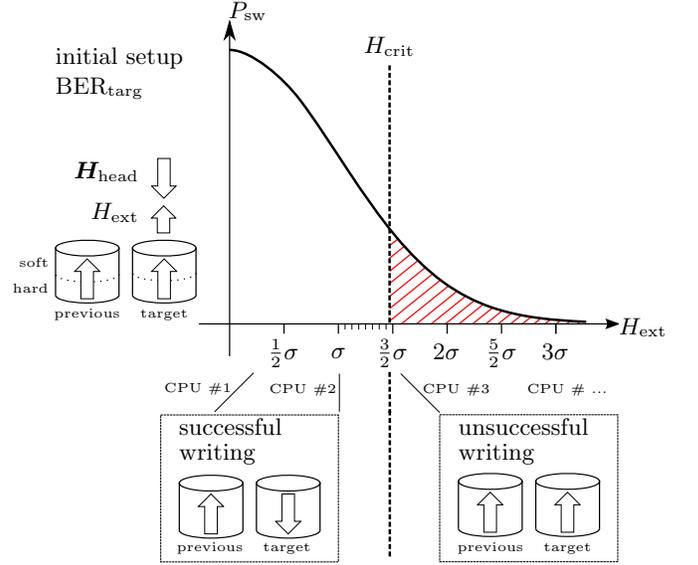
\begin{figure}
\begingroup%
  \makeatletter%
  \providecommand\color[2][]{%
    \errmessage{(Inkscape) Color is used for the text in Inkscape, but the package 'color.sty' is not loaded}%
    \renewcommand\color[2][]{}%
  }%
  \providecommand\transparent[1]{%
    \errmessage{(Inkscape) Transparency is used (non-zero) for the text in Inkscape, but the package 'transparent.sty' is not loaded}%
    \renewcommand\transparent[1]{}%
  }%
  \providecommand\rotatebox[2]{#2}%
  \ifx\svgwidth\undefined%
    \setlength{\unitlength}{219.325bp}%
    \ifx\svgscale\undefined%
      \relax%
    \else%
      \setlength{\unitlength}{\unitlength * \real{\svgscale}}%
    \fi%
  \else%
    \setlength{\unitlength}{\svgwidth}%
  \fi%
  \global\let\svgwidth\undefined%
  \global\let\svgscale\undefined%
  \makeatother%
  \begin{picture}(1,0.93423002)%
    \put(0,0){\includegraphics[width=\unitlength]{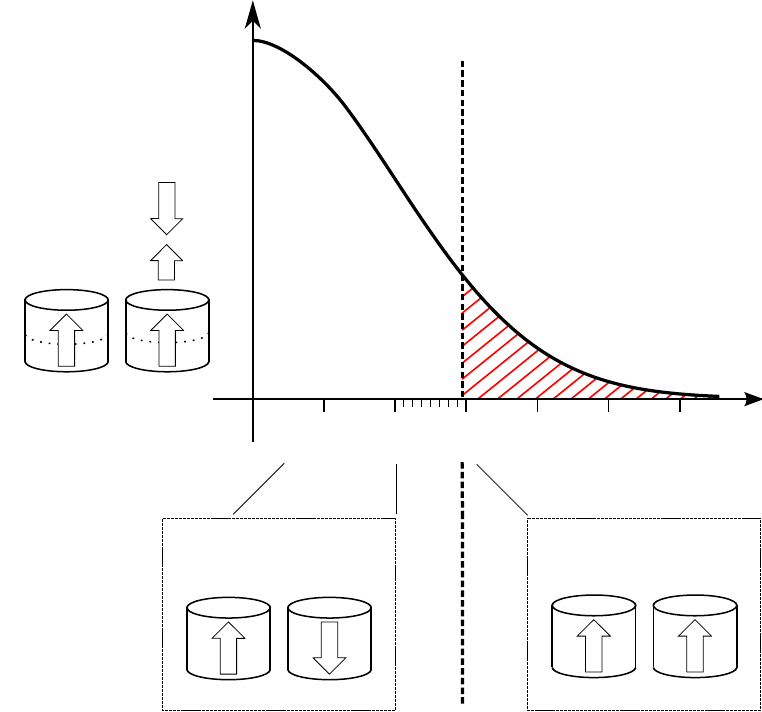}}%
    \put(0.33004555,0.93636292){\color[rgb]{0,0,0}\makebox(0,0)[lb]{\smash{$P_{\mathrm{sw}}$}}}%
    \put(0.60682495,0.87581148){\color[rgb]{0,0,0}\makebox(0,0)[b]{\smash{$H_{\mathrm{crit}}$}}}%
    \put(0.42435459,0.34959223){\color[rgb]{0,0,0}\makebox(0,0)[b]{\smash{$\frac{1}{2}\sigma$}}}%
    \put(0.61121606,0.34959223){\color[rgb]{0,0,0}\makebox(0,0)[b]{\smash{$\frac{3}{2}\sigma$}}}%
    \put(0.79861479,0.34977479){\color[rgb]{0,0,0}\makebox(0,0)[b]{\smash{$\frac{5}{2}\sigma$}}}%
    \put(0.70464678,0.34959223){\color[rgb]{0,0,0}\makebox(0,0)[b]{\smash{$2\sigma$}}}%
    \put(0.89150828,0.34959223){\color[rgb]{0,0,0}\makebox(0,0)[b]{\smash{$3\sigma$}}}%
    \put(1.0008446,0.3923359){\color[rgb]{0,0,0}\makebox(0,0)[lb]{\smash{$H_{\mathrm{ext}}$}}}%
    \put(0.51884842,0.35065529){\color[rgb]{0,0,0}\makebox(0,0)[b]{\smash{$\sigma$}}}%
    \put(0.0872928,0.42088857){\color[rgb]{0,0,0}\makebox(0,0)[b]{\smash{{\tiny previous}}}}%
    \put(0.21956503,0.42088857){\color[rgb]{0,0,0}\makebox(0,0)[b]{\smash{{\tiny target}}}}%
    \put(0.02085562,0.50809518){\color[rgb]{0,0,0}\makebox(0,0)[rb]{\smash{{\tiny soft}}}}%
    \put(0.02089891,0.46350904){\color[rgb]{0,0,0}\makebox(0,0)[rb]{\smash{{\tiny hard}}}}%
    \put(0.1761694,0.58977577){\color[rgb]{0,0,0}\makebox(0,0)[rb]{\smash{$H_{\mathrm{ext}}$}}}%
    \put(0.17642236,0.66026371){\color[rgb]{0,0,0}\makebox(0,0)[rb]{\smash{$\bm{H}_{\mathrm{head}}$}}}%
    \put(0.03291277,0.80628339){\color[rgb]{0,0,0}\makebox(0,0)[lb]{\smash{$\mathrm{BER}_{\mathrm{targ}}$}}}%
    \put(0.03291277,0.85736209){\color[rgb]{0,0,0}\makebox(0,0)[lb]{\smash{initial setup }}}%
    \put(0.29826182,0.01937296){\color[rgb]{0,0,0}\makebox(0,0)[b]{\smash{{\tiny previous}}}}%
    \put(0.43053405,0.01937296){\color[rgb]{0,0,0}\makebox(0,0)[b]{\smash{{\tiny target}}}}%
    \put(0.24552768,0.21930924){\color[rgb]{0,0,0}\makebox(0,0)[lb]{\smash{successful}}}%
    \put(0.24553079,0.17688415){\color[rgb]{0,0,0}\makebox(0,0)[lb]{\smash{writing}}}%
    \put(0.78034585,0.01937289){\color[rgb]{0,0,0}\makebox(0,0)[b]{\smash{{\tiny previous}}}}%
    \put(0.91261803,0.01937289){\color[rgb]{0,0,0}\makebox(0,0)[b]{\smash{{\tiny target}}}}%
    \put(0.72515265,0.21928297){\color[rgb]{0,0,0}\makebox(0,0)[lb]{\smash{unsuccessful}}}%
    \put(0.72511853,0.17689462){\color[rgb]{0,0,0}\makebox(0,0)[lb]{\smash{writing}}}%
    \put(0.33548629,0.2924527){\color[rgb]{0,0,0}\makebox(0,0)[rb]{\smash{{\tiny CPU \#1}}}}%
    \put(0.51577157,0.29166503){\color[rgb]{0,0,0}\makebox(0,0)[rb]{\smash{{\tiny CPU \#2}}}}%
    \put(0.66372878,0.29103054){\color[rgb]{0,0,0}\makebox(0,0)[lb]{\smash{{\tiny CPU \#3}}}}%
    \put(0.98140507,0.29166503){\color[rgb]{0,0,0}\makebox(0,0)[rb]{\smash{{\tiny CPU \# ...}}}}%
  \end{picture}%
\endgroup%
\caption{Calculation of target bit error rate with an assumed switching field distribution $P_{\mathrm{sw}}$ with $\sigma=0.015H_{\mathrm{A}}$ . $H_{\mathrm{crit,targ}}$ is the anti-parallel aligned external field $H_{\mathrm{ext}}$ which is needed to prevent switching of the target bit. Calculation of the previous bit error rate is analogous. We search for $H_{\mathrm{crit,prev}}$ which is the parallel applied field $H_{\mathrm{ext}}$ where keeping the previous bit from getting rewritten with $H_{\mathrm{head}}$ isn't achievable any more.}
\label{fig:gauss}
\end{figure}

Secondly we perform micromagnetic recording simulations. During a single micromagnetic simulation we locally apply the previously calculated write head field profile $\bm{H}_{\mathrm{head}}$ shifted accordingly with the down-track and cross-track offset. The interaction from the media onto the pole and underlayer is not included into the precomputed write field calculation. Additionaly, an external field $H_{\mathrm{ext}}$ is globally applied, which functions as a negative influence on the write head field. In other words this field works against the recording head field. Similarly, we can imagine that variations of the anisotropy field and the magnetostatic interaction field hinder the switching in a given head field. The additional external field mimics this situation. This additionally applied field $H_{\mathrm{ext}}$ is anti-parallel or parallel aligned to the head field for calculating the target bit or previous bit error rate, respectively. The purpose of these simulations is to find the critical external field strength $H_{\mathrm{crit,targ}}$ and $H_{\mathrm{crit,prev}}$ where switching of the target bit or not switching the previous bit isn't achievable any more. The recording simulations can be done in parallel where one has different $H_{\mathrm{ext}}$ applied. Each one of 12 CPUs perform a recording simulation where only $H_{\mathrm{ext}}$ is taken from the set $\{0,\cdots,4\sigma\}$ (see FIG.~\ref{fig:gauss}) and the critical field, in our case, has always been found within this range.

After determining maximal allowed fields $H_{\mathrm{crit,targ}}$ and $H_{\mathrm{crit,prev}}$ we are able to obtain both bit error rates by integrating over the tail of the Gaussian function with the critical fields as lower bound~\citep{dong2012}, as shown in the following equations \eqref{eq:ber} and \eqref{eq:ber-02}.
\begin{align}
\mathrm{BER}_{\mathrm{targ}}&=\frac{1}{2}\left[1-\mathrm{erf}\left(\frac{H_{\mathrm{crit,targ}}}{\sqrt{2\sigma^2}}\right)\right]
\label{eq:ber}
\\
\mathrm{BER}_{\mathrm{prev}}&=\frac{1}{2}\left[1-\mathrm{erf}\left(\frac{H_{\mathrm{crit,prev}}}{\sqrt{2\sigma^2}}\right)\right]
\label{eq:ber-02}
\end{align}
The part of the evaluation script, which searches for the critical field $H_{\mathrm{crit}}$ is able to refine the search range multiple times.
In this work the search pattern has been refined twice with an initial search discretization of half the estimated switching field distribution $\sigma$ (illustrated with long and short tic marks at $H_{\mathrm{ext}}$-Axis in FIG. ~\ref{fig:gauss}). The switching field distribution is assumed to control the anisotropy variation\citep{richter2006a}.
For these recording simulations the write field moves with $\SI{13.5}{\meter/\second}$ in down track direction and is scaled with a rise time of $\SI{0.1}{\nano\second}$ and reverses within a period of $\SI{13.5}{\nano\second}$.

For calculating the thermally induced adjacent track erasure error $\mathrm{BER}_{\mathrm{adj}}$, we place a single bit into the fringing field of the head (see adjacent bit position in FIG.~\ref{fig:best-profiles}). While influenced by this field we compute the energy barrier $E_\mathrm{B}$ for switching micromagnetically with the nudged elastic band method. The highest field and the lowest barrier occur at the down track location which is \SI{2}{nm} from the trailing edge in flight direction. At this fixed location the energy barriers for computing the adjacent track error was always evaluated.
From the energy barrier we calculate the life time of the bit $\tau=f_{0}^{-1}\mathrm{exp}\left(E_\mathrm{B}/k_{\mathrm{B}}T\right)$ with an attempt frequency of $f_0=\SI{130}{GHz}$. The attempt frequency was calculated for similar intrinsic media property composition\citep{Dean2008} by micromagnetically solving the stochastic Landau-Lifshitz-Gilbert equation\citep{Tsiantos2003}. We assume a field exposure time during writing $t_{\mathrm{write}}=\SI{1}{\nano\second}$. Then the number of passes before erasure is $\tau/t_{\mathrm{write}}$ and 
\begin{equation}
\mathrm{BER}_{\mathrm{adj}}=\frac{t_{\mathrm{write}}}{\tau}\mathrm{.}
\label{eq:ber-03}
\end{equation}
The total bit error rate is the sum of the above described error rates $\mathrm{BER}_{\mathrm{tot}}=\mathrm{BER}_{\mathrm{targ}}+\mathrm{BER}_{\mathrm{prev}}+\mathrm{BER}_{\mathrm{adj}}$.  
This bit error rate gives the probability of an error during a single write event.
The simulations for each kind of error rate are resource wise expensive, hence they are calculated sequentially to be able to abort the remaining {\textit{model evaluation}} as soon as one error rate doesn't meet the requirement.
For example if the current evaluated write head design can't reverse the target bit with its write field, there is no reason to investigate how it would perform at the other error rates. Similarly if the head design is performing well at target bit error rate but always re-writes the previous bit too, there is no need to compute the adjacent error rate any more.
\section{\label{sec:results}Results and Discussion}

\begingroup
\squeezetable
\renewcommand\arraystretch{1.2}
\begin{table*}{}
\centering
\begin{ruledtabular}
\begin{tabular}{c | c c c | c | r r r r r r r r r r r}
& & & & & trailing & trailing & side & side & pole & tip &  &  &  & down & cross \\
& & & & $E_{\mathrm{B,0}}$  & shield & edge & shield & edge & tip & taper & shield &  &  & track & track \\
& & & & $\left(k_{\mathrm{B}}T\right)$ & gap & angle & gap & angle & width & angle & thickness & $K_{\mathrm{1,hard}}$ & $A_{\mathrm{ex,int}}$ & offset & offset \\ 
$\mathrm{BER}_{\mathrm{tot}}$ & $\mathrm{BER}_{\mathrm{targ}}$ & $\mathrm{BER}_{\mathrm{prev}}$ & $\mathrm{BER}_{\mathrm{adj}}$ & $T=\SI{300}{K}$ & $\left(\mathrm{\si{\nano\meter}}\right)$ & $\left(\mathrm{\si{\degree}}\right)$ & $\left(\mathrm{\si{\nano\meter}}\right)$ & $\left(\mathrm{\si{\degree}}\right)$ & $\left(\mathrm{\si{\nano\meter}}\right)$ & $\left(\mathrm{\si{\degree}}\right)$ & $\left(\mathrm{\si{\nano\meter}}\right)$ & $\left(\mathrm{\si{\kilo\joule/\cubic\meter}}\right)$ & $\left(\mathrm{\si{\pico\joule/\meter}}\right)$ & $\left(\mathrm{\si{\nano\meter}}\right)$ & $\left(\mathrm{\si{\nano\meter}}\right)$ \\
\hline
\textbf{centered} &&&&&&&&&&&&&\\
$1.3\mathrm{x}10^{-2}$ & $5.9\mathrm{x}10^{-3}$ & $3.8\mathrm{x}10^{-3}$ & $3.4\mathrm{x}10^{-3}$ & 140 & 15 & 31 & 12 & 45 & 5.3 & 75 & 12 & 660 & 2.6 & 5.4 & 0.0\\
$2.9\mathrm{x}10^{-3}$ & $2.8\mathrm{x}10^{-3}$ & $3.2\mathrm{x}10^{-5}$ & $3.6\mathrm{x}10^{-6}$ & 110 & 14 & 38 & 14 & 18 & 18 & 75 & 9.9 & 800 & 1.3 & 2.9 & 0.0\\
\hline
\textbf{staggered} &&&&&&&&&&&&&\\
$2.1\mathrm{x}10^{-2}$ & $2.1\mathrm{x}10^{-3}$ & $1.9\mathrm{x}10^{-2}$ & $5.3\mathrm{x}10^{-17}$ & 110 & 13 & 24 & 11 & 31 & 12 & 75 & 12 & 620 & 1.1 & 3.6 & 0.0\\
$1.8\mathrm{x}10^{-2}$ & $8.9\mathrm{x}10^{-3}$ & $8.9\mathrm{x}10^{-3}$ & $2.8\mathrm{x}10^{-18}$ & 110 & 14 & 22 & 12 & 31 & 12 & 75 & 12 & 660 & 1.2 & 3.8 & 0.0\\
\hline
\textbf{shingled} &&&&&&&&&&&&&\\
$8.4\mathrm{x}10^{-8}$ & $4.2\mathrm{x}10^{-8}$ & $4.2\mathrm{x}10^{-8}$ & $3.2\mathrm{x}10^{-15}$ & 130 & 10 & 19 & 11 & 31 & 80 & 75 & 50 & 730 & 5.3 & 7.2 & 5.6 \\
$1.5\mathrm{x}10^{-8}$ & $9.9\mathrm{x}10^{-10}$ & $9.9\mathrm{x}10^{-10}$ & $1.3\mathrm{x}10^{-8}$ & 140 & 13 & 27 & 16 & 33 & 80 & 75 & 50 & 810 & 10 & 3.3 & 5.1
\end{tabular}
\caption{\label{tab:final-results}Bit error rates and parameters for two best designs for the three writing schemes: centered writing, staggered writing and shingled writing. All values are rounded to two significant digits.}
\end{ruledtabular}
\end{table*}
\endgroup

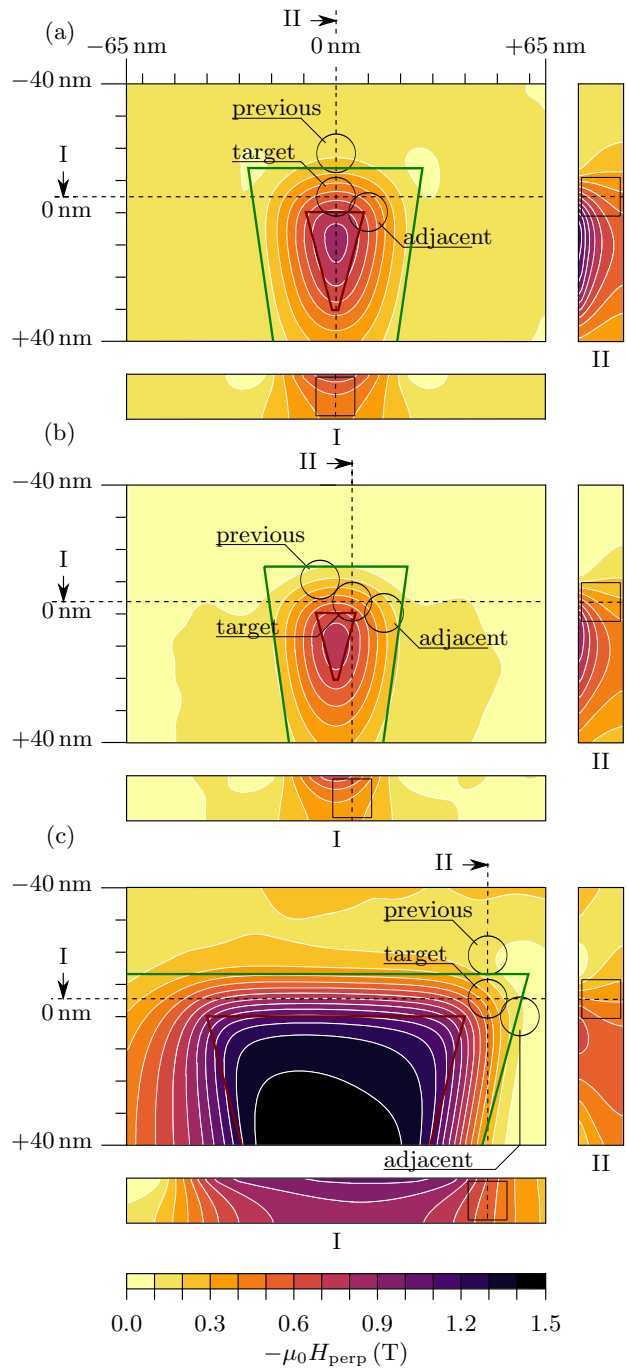
\begin{figure}
\begingroup%
  \makeatletter%
  \providecommand\color[2][]{%
    \errmessage{(Inkscape) Color is used for the text in Inkscape, but the package 'color.sty' is not loaded}%
    \renewcommand\color[2][]{}%
  }%
  \providecommand\transparent[1]{%
    \errmessage{(Inkscape) Transparency is used (non-zero) for the text in Inkscape, but the package 'transparent.sty' is not loaded}%
    \renewcommand\transparent[1]{}%
  }%
  \providecommand\rotatebox[2]{#2}%
  \ifx\svgwidth\undefined%
    \setlength{\unitlength}{216.11520996bp}%
    \ifx\svgscale\undefined%
      \relax%
    \else%
      \setlength{\unitlength}{\unitlength * \real{\svgscale}}%
    \fi%
  \else%
    \setlength{\unitlength}{\svgwidth}%
  \fi%
  \global\let\svgwidth\undefined%
  \global\let\svgscale\undefined%
  \makeatother%
  \begin{picture}(1,2.36084563)%
    \put(0,0){\includegraphics[width=\unitlength]{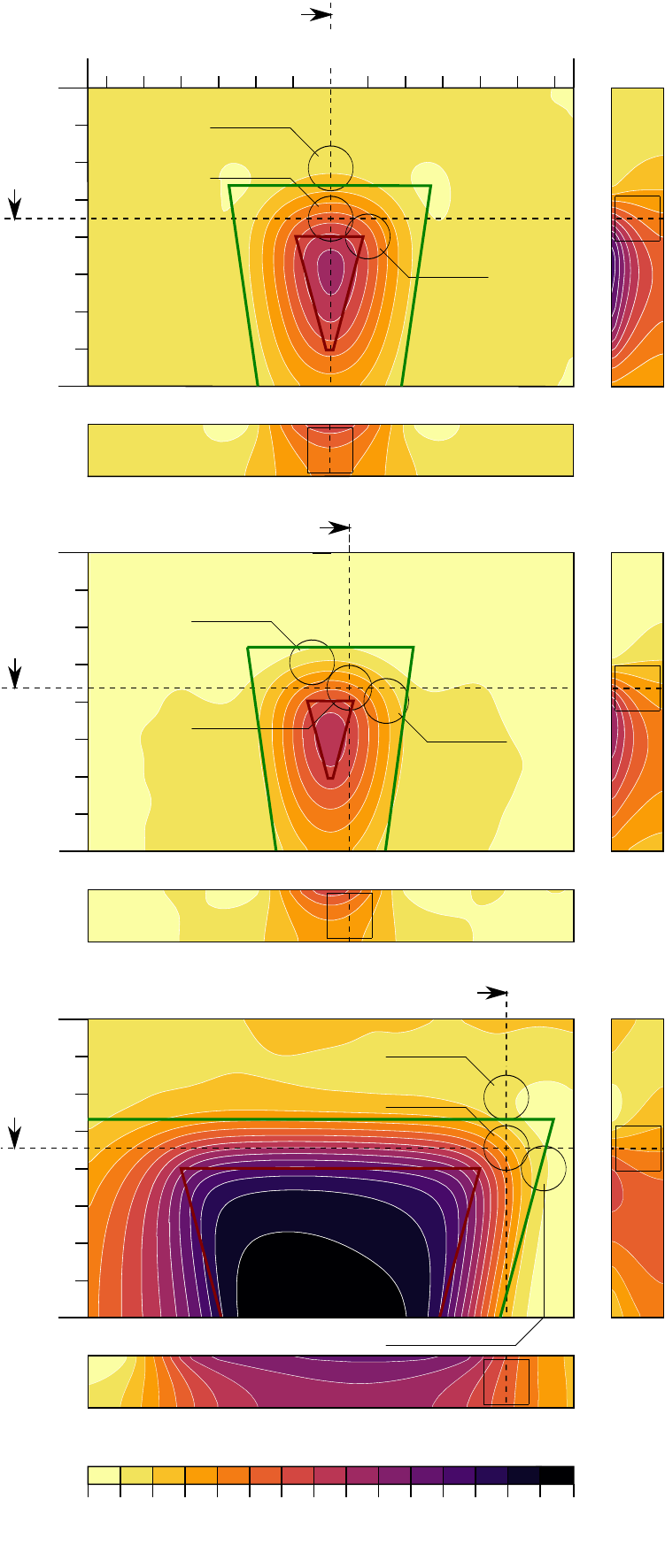}}%
    \put(0.95984487,1.73402227){\color[rgb]{0,0,0}\makebox(0,0)[b]{\smash{\textrm{II}}}}%
    \put(0.49775048,2.28275664){\color[rgb]{0,0,0}\makebox(0,0)[b]{\smash{$\SI{0}{nm}$}}}%
    \put(0.07190702,1.99706466){\color[rgb]{0,0,0}\makebox(0,0)[rb]{\smash{$\SI{0}{nm}$}}}%
    \put(0.13213242,2.28275664){\color[rgb]{0,0,0}\makebox(0,0)[b]{\smash{$-\SI{65}{nm}$}}}%
    \put(0.86323841,2.28275664){\color[rgb]{0,0,0}\makebox(0,0)[b]{\smash{$+\SI{65}{nm}$}}}%
    \put(0.07281799,2.22294064){\color[rgb]{0,0,0}\makebox(0,0)[rb]{\smash{$-\SI{40}{nm}$}}}%
    \put(0.07319395,1.77371916){\color[rgb]{0,0,0}\makebox(0,0)[rb]{\smash{$+\SI{40}{nm}$}}}%
    \put(0.02238193,2.09180579){\color[rgb]{0,0,0}\makebox(0,0)[b]{\smash{\textrm{I}}}}%
    \put(0.49905187,1.598535){\color[rgb]{0,0,0}\makebox(0,0)[b]{\smash{\textrm{I}}}}%
    \put(0.04392715,2.30660731){\color[rgb]{0,0,0}\makebox(0,0)[rb]{\smash{(a)}}}%
    \put(0.04431757,1.60649878){\color[rgb]{0,0,0}\makebox(0,0)[rb]{\smash{(b)}}}%
    \put(0.04392715,0.9037215){\color[rgb]{0,0,0}\makebox(0,0)[rb]{\smash{(c)}}}%
    \put(0.4372648,2.32967624){\color[rgb]{0,0,0}\makebox(0,0)[rb]{\smash{\textrm{II}}}}%
    \put(0.49905187,0.89737){\color[rgb]{0,0,0}\makebox(0,0)[b]{\smash{\textrm{I}}}}%
    \put(0.95958459,1.03381614){\color[rgb]{0,0,0}\makebox(0,0)[b]{\smash{\textrm{II}}}}%
    \put(0.07216729,1.29695523){\color[rgb]{0,0,0}\makebox(0,0)[rb]{\smash{$\SI{0}{nm}$}}}%
    \put(0.07319395,1.52283211){\color[rgb]{0,0,0}\makebox(0,0)[rb]{\smash{$-\SI{40}{nm}$}}}%
    \put(0.07358436,1.07361063){\color[rgb]{0,0,0}\makebox(0,0)[rb]{\smash{$+\SI{40}{nm}$}}}%
    \put(0.02264221,1.38607147){\color[rgb]{0,0,0}\makebox(0,0)[b]{\smash{\textrm{I}}}}%
    \put(0.46512899,1.5573379){\color[rgb]{0,0,0}\makebox(0,0)[rb]{\smash{\textrm{II}}}}%
    \put(0.13252283,0.05292385){\color[rgb]{0,0,0}\makebox(0,0)[b]{\smash{0.0}}}%
    \put(0.27837964,0.05292385){\color[rgb]{0,0,0}\makebox(0,0)[b]{\smash{0.3}}}%
    \put(0.42412077,0.05302056){\color[rgb]{0,0,0}\makebox(0,0)[b]{\smash{0.6}}}%
    \put(0.56984744,0.05292385){\color[rgb]{0,0,0}\makebox(0,0)[b]{\smash{0.9}}}%
    \put(0.71557411,0.05288409){\color[rgb]{0,0,0}\makebox(0,0)[b]{\smash{1.2}}}%
    \put(0.86312273,0.05295639){\color[rgb]{0,0,0}\makebox(0,0)[b]{\smash{1.5}}}%
    \put(0.49788062,0.00131383){\color[rgb]{0,0,0}\makebox(0,0)[b]{\smash{$-\mu_0H_{\mathrm{perp}} \left(\mathrm{\si{\tesla}}\right)$}}}%
    \put(0.07177688,0.59417885){\color[rgb]{0,0,0}\makebox(0,0)[rb]{\smash{$\SI{0}{nm}$}}}%
    \put(0.07281799,0.82006296){\color[rgb]{0,0,0}\makebox(0,0)[rb]{\smash{$-\SI{40}{nm}$}}}%
    \put(0.07319395,0.37083426){\color[rgb]{0,0,0}\makebox(0,0)[rb]{\smash{$+\SI{40}{nm}$}}}%
    \put(0.02238193,0.69332935){\color[rgb]{0,0,0}\makebox(0,0)[b]{\smash{\textrm{I}}}}%
    \put(0.49905187,0.19530306){\color[rgb]{0,0,0}\makebox(0,0)[b]{\smash{\textrm{I}}}}%
    \put(0.95984487,0.33161183){\color[rgb]{0,0,0}\makebox(0,0)[b]{\smash{\textrm{II}}}}%
    \put(0.70203967,0.85698808){\color[rgb]{0,0,0}\makebox(0,0)[rb]{\smash{\textrm{II}}}}%
    \put(0.58106831,0.69841561){\color[rgb]{0,0,0}\makebox(0,0)[lb]{\smash{target}}}%
    \put(0.58080803,0.77421611){\color[rgb]{0,0,0}\makebox(0,0)[lb]{\smash{previous}}}%
    \put(0.58119845,0.33923671){\color[rgb]{0,0,0}\makebox(0,0)[lb]{\smash{adjacent}}}%
    \put(0.28844372,1.42977465){\color[rgb]{0,0,0}\makebox(0,0)[lb]{\smash{previous}}}%
    \put(0.28883414,1.26885697){\color[rgb]{0,0,0}\makebox(0,0)[lb]{\smash{target}}}%
    \put(0.64258064,1.24907676){\color[rgb]{0,0,0}\makebox(0,0)[lb]{\smash{adjacent}}}%
    \put(0.31655373,2.17369932){\color[rgb]{0,0,0}\makebox(0,0)[lb]{\smash{previous}}}%
    \put(0.31655373,2.09789973){\color[rgb]{0,0,0}\makebox(0,0)[lb]{\smash{target}}}%
    \put(0.61473091,1.94857345){\color[rgb]{0,0,0}\makebox(0,0)[lb]{\smash{adjacent}}}%
  \end{picture}%
\endgroup%
\caption{(color-online) Perpendicular field profile of the optimized design for (a) centered writing (b) staggered writing and (c) shingled writing. Each field profile is shown in three perspectives: top view $\SI{9}{nm}$ below the air bearing surface, (\textrm{I}) cross track view along down track offset and (\textrm{II}) down track view along cross track offset. Each top view shows additionally the contours of the best head's air bearing surface and shield. In the top views positions of target, previous and adjacent bits are shown in circles. The cross sections always show the position of the target bit only. Note that due to worst case location placement of the adjacent bit it might overlap with the other bits.}
\label{fig:best-profiles}
\end{figure}

The use of several design parameters during an optimization may result in an objective function landscape which has more than one local minimum with comparable objective function values. Thus more than one design with similar performance is possible. As mentioned before it is recommended to use at least ten times the amount of design parameters as initial training points. By this method we hope to find the most promising designs. 
The two best designs for each writing scheme are summarized in TABLE~\ref{tab:final-results}. The field profiles of each writing scheme's best design are shown in FIG.~\ref{fig:best-profiles}. Each field profile is shown from the top, cross- and down-track direction. The top view shows an $xy$-plane slice of the field profile $\SI{9}{nm}$ below the air bearing surface, which corresponds to the middle of the media's soft phase. The cross- and down-track slices are taken depending on the location of the target bit which is furthermore dependent on optimized down- and cross-track offset. The location of target, previous and adjacent bit are illustrated as labelled contours in the FIG.~\ref{fig:best-profiles}.

For \textbf{centered writing} (FIG.~\ref{fig:best-profiles}a) the number of free parameters was 9 (described in Section~\ref{sec:design-params}). After 46 function evaluations the total bit error rate $\mathrm{BER}_{\mathrm{tot}}$ was minimized to $2.9\mathrm{x}10^{-3}$. With this scheme it is most difficult to correctly write the target bit. The target bit error rate $\mathrm{BER}_{\mathrm{targ}}$ dominates the overall error rate.

For \textbf{staggered writing} (FIG.~\ref{fig:best-profiles}b) the number of free parameters was 9. After 94 function evaluations the total bit error rate $\mathrm{BER}_{\mathrm{tot}}$ was minimized to $1.8\mathrm{x}10^{-2}$. The short down track distance between bits on the target track increases the error. Write errors for the target bit occur at the same rate as rewriting previously written bits, therefore $\mathrm{BER}_{\mathrm{targ}}$ and $\mathrm{BER}_{\mathrm{prev}}$ are equal $8.9\mathrm{x}10^{-3}$. 

For \textbf{shingled writing} (FIG.~\ref{fig:best-profiles}c) the number of free parameters was 8 (described in Section~\ref{sec:design-params}). Total bit error rate was $1.5\mathrm{x}10^{-8}$ after 62 function evaluations. The overall bit error rate is dominated by the probability of thermally switching an island on the adjacent track. A similar performance, with slightly changed trailing edge angle and side shield gap but with interfacial exchange coupling $A_{\mathrm{ex,int}}$ halved, was reached after 68 function evaluations with a total bit error rate of $\mathrm{BER}_{\mathrm{tot}}=8.4\mathrm{x}10^{-8}$.

Additionally we calculated the zero field energy barriers $E_{\mathrm{B,0}}$ (shown in TABLE~\ref{tab:final-results}) with the nudged elastic band method, for each best exchange coupled-composite media design. The zero field energy barrier of the two best centered writing designs are $\mathrm{140} k_{\mathrm{B}}T$ and $\mathrm{110} k_{\mathrm{B}}T$ with $T=\SI{300}{K}$. For staggered writing both barriers are identical at $\mathrm{110} k_{\mathrm{B}}T$ and for shingled writing we calculated zero field energy barriers of $\mathrm{140} k_{\mathrm{B}}T$ and $\mathrm{130} k_{\mathrm{B}}T$.
\section{\label{sec:conclusion}Conclusion}
We developed an algorithm for the joint optimization of writer and media properties in order to find optimal design parameters for bit patterned media recording. The evaluation of the error rate is fully based on micro magnetic simulations which not only take into account the dynamics on track error but also the thermally induced adjacent track erasure. The computational framework of this paper can be scaled to span large compute clusters, thus enabling optimum solutions for technologically important problems that have rich design spaces.

The optimization runs show that shingled writing clearly outperforms all the other writing schemes for high areal densities. 

In centered writing the constraint on pole dimension required to avoid adjacent track erasure on both sides limits the maximum head field which introduces errors for writing the target bit. In staggered writing the effective down-track bit separation is lower by one half as compared to centered or shingled writing. Here our results show high bit error rates caused by back switching the previous bit.

In the design parameters found for centered and staggered writing the perpendicular write field is well below $\SI{1}{T}$. Decreasing the air bearing surface to media spacing from $\SI{6}{nm}$ as used in our simulations, will be essential to achieve error rates below $10^{-3}$ with centered writing on $\SI{4}{Tb/in^2}$ exchange coupled composite bit patterned media. On the other hand write error rates in the range of $10^{-8}$ were achieved for shingled writing.

\begin{acknowledgments}
We acknowledge the financial support from ASTC / IDEMA, the Austrian Science Fund (FWF Project Nr.: I821) and the Vienna Science and Technology Fund (WWTF Grant No. MA14-044).
\end{acknowledgments}
\bibliography{biblio}
\end{document}